\documentclass[aps,showkeys,twocolumn,showpacs,preprintnumbers,amsmath,amssymb,superscriptaddress,floatfix,nofootinbib]{revtex4}
\usepackage{graphicx,color,dcolumn,booktabs,bm}
\usepackage{longtable,lscape}
\usepackage{txfonts}
\usepackage{overpic}
\usepackage{epsfig}
\usepackage{amssymb}
\usepackage{rotating}
\usepackage{epstopdf}
\usepackage{indentfirst}
\usepackage{feynmf}   
\usepackage{slashed}  
\usepackage{cases}
\usepackage{color}
\usepackage{multirow}
\usepackage{graphicx,color,dcolumn,booktabs,bm}
\usepackage{cases}
\usepackage{array}

\begin{document}

\title{Strong decays of the newly observed narrow $\Omega_b$ structures }
\author{Wei Liang}
\affiliation{  Department
of Physics, Hunan Normal University,  Changsha 410081, China }

\affiliation{ Synergetic Innovation
Center for Quantum Effects and Applications (SICQEA), Changsha 410081,China}

\affiliation{  Key Laboratory of
Low-Dimensional Quantum Structures and Quantum Control of Ministry
of Education, Changsha 410081, China}

\author{Qi-Fang L\"u \footnote{Corresponding author} } \email{lvqifang@hunnu.edu.cn} %
\affiliation{  Department
of Physics, Hunan Normal University,  Changsha 410081, China }

\affiliation{ Synergetic Innovation
Center for Quantum Effects and Applications (SICQEA), Changsha 410081,China}

\affiliation{  Key Laboratory of
Low-Dimensional Quantum Structures and Quantum Control of Ministry
of Education, Changsha 410081, China}

\begin{abstract}

Motivated by the newly observed narrow structures $\Omega_b(6316)^-$, $\Omega_b(6330)^-$, $\Omega_b(6340)^-$, and $\Omega_b(6350)^-$ in the $\Xi_b^0 K^-$ mass spectrum, we investigate the strong decays of the low-lying $\Omega_b$ states within the $^3P_0$ model systematically. According to their masses and decay widths, the observed $\Omega_b(6316)^-$, $\Omega_b(6330)^-$, $\Omega_b(6340)^-$, and $\Omega_b(6350)^-$ resonances can be reasonably assigned as the $\lambda-$mode $\Omega_b(1P)$ states with $J^P=1/2^-, 3/2^-, 3/2^-$, and $5/2^-$. Meanwhile, the remaining  $P-$wave state with $J^P=1/2^-$ should have a rather broad width, which can hardly be observed by experiments. For the $\Omega_b(2S)$ and $\Omega_b(1D)$ states, our predictions show that these states have relatively narrow total widths and mainly decay into the $\Xi_b \bar K$, $\Xi_b^\prime \bar K$ and $\Xi_b^{\prime*} \bar K$ final states. These abundant theoretical predictions may be valuable for searching more excited $\Omega_b$ states in future experiments.

\end{abstract}

\keywords{Low-lying $\Omega_b$ states; Strong decays; $^3P_0$ model}

\maketitle

\section{Introduction}

\begin{figure}[!htbp]
\includegraphics[scale=0.55]{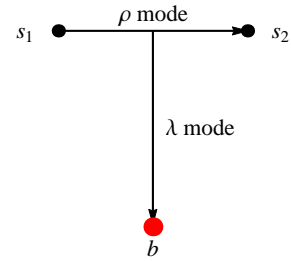}
\vspace{0.0cm} \caption{The $\Omega_b$ system with $\lambda-$ or $\rho-$mode excitation. The $s_1$ and $s_2$ stand for the strange quarks, and $b$ corresponds to the bottom quark.}\label{jacobi}
\end{figure}

Two kinds of excitations, the $\rho$-mode and $\lambda$-mode, exist for the $\Omega_b$ family. The $\rho$-mode one is the excitation between two strange quarks, while the $\lambda$-mode case is the excitation between the strange quark subsystem and bottom quark. Such a structure is illustrated in Fig.~\ref{jacobi}. For this system, the $\lambda$-mode is more easily excited due to its heavy reduced mass. To understand the internal structures of these states and establish the low-lying $\Omega_b$ spectrum, experimental and theoretical efforts on their strong decay behaviors are urgently needed.

Although the constituent quark models have predicted plenty of heavy baryons for a long time~\cite{Chen:2016spr,Cheng:2015iom,Crede:2013sze,Klempt:2009pi}, the experimental information on the $\Omega_c$ and $\Omega_b$ states was scarce in the past years~\cite{Tanabashi:2018oc}. Before 2017, there were only three $\Omega_c$ and $\Omega_b$ ground states in experiments: $\Omega_c (2695)$, $\Omega_c (2770)$ and $\Omega_b (6046)$ , while the ground state $\Omega_b^*$ was missing. In 2017, the LHCb Collaboration observed five narrow resonances $\Omega_c (3000)$, $\Omega_c (3050)$, $\Omega_c (3066)$, $\Omega_c (3090)$, and $\Omega_c (3119)$ in the $\Xi^+_c K^-$ channel~\cite{Aaij:2017nav}. Meanwhile, the evidence of a relatively broad signal $\Omega_c (3188)$ was also reported~\cite{Aaij:2017nav}. Subsequently, the Belle Collaboration confirmed most of them except for the $\Omega_c (3119)$ structure~\cite{Yelton:2017qxg}.

Very recently, the LHCb Collaboration reported four narrow peaks $\Omega_b(6316)^-$, $\Omega_b(6330)^-$, $\Omega_b(6340)^-$, and $\Omega_b(6350)^-$ in the $\Xi_b^0 K^-$ mass spectrum~\cite{Aaij:2020cex}. Their measured masses and decay widths are listed as follows:
\begin{eqnarray}
m[\Omega_b(6316)^-] = 6315.64\pm0.31~\pm0.07\pm0.50~\rm{MeV},
\end{eqnarray}
\begin{eqnarray}
\Gamma[\Omega_b(6316)^-] < 2.8~\rm{MeV},
\end{eqnarray}
\begin{eqnarray}
m[\Omega_b(6330)^-] = 6330.30\pm0.28\pm0.07\pm0.50~\rm{MeV},
\end{eqnarray}
\begin{eqnarray}
\Gamma[\Omega_b(6330)^-] <  3.1~\rm{MeV},
\end{eqnarray}
\begin{eqnarray}
m[\Omega_b(6340)^-] = 6339.71\pm0.26~\pm0.05\pm0.50~\rm{MeV},
\end{eqnarray}
\begin{eqnarray}
\Gamma[\Omega_b(6340)^-] < 1.5~\rm{MeV},
\end{eqnarray}
\begin{eqnarray}
m[\Omega_b(6350)^-] = 6349.88\pm0.35\pm0.05\pm0.50~\rm{MeV},
\end{eqnarray}
\begin{eqnarray}
\Gamma[\Omega_b(6350)^-] = 1.4^{+1.0}_{-0.8}\pm0.1~\rm{MeV},
\end{eqnarray}
where only the upper limits are determined for the three lower mass states. Compared with the masses predicted by the constituent quark models~\cite{Ebert:2011kk,Roberts:2007ni,Ebert:2007nw,Garcilazo:2007eh,Yoshida:2015tia,Thakkar:2016dna,Shah:2018daf,Qin:2019hgk,Chen:2018vuc}, these four resonances are good candidates of the $\lambda$-mode $\Omega_b(1P)$ states.

The observations of the low-lying $\Omega_c$ resonances have made an important progress towards a better understanding of the heavy baryon spectrum and aroused widespread interests in hadron physics. Lots of conventional and exotic interpretations on these low-lying states have been done, especially for their strong decays~\cite{Chen:2017sci,Karliner:2017kfm,Wang:2017hej,Wang:2017vnc,Padmanath:2017lng,Cheng:2017ove,Huang:2017dwn,Wang:2017zjw,Zhao:2017fov,
Chen:2017gnu,Kim:2017jpx,Agaev:2017lip,Wang:2017xam,Wang:2017kfr,Debastiani:2017ewu,Yao:2018jmc,Santopinto:2018ljf,
Yang:2017rpg,An:2017lwg,Montana:2017kjw,Wang:2017smo,Huang:2018wgr,Debastiani:2018adr,Wang:2018alb,Agaev:2017ywp}. However, the conclusions from various theoretical works with different models and parameters are not consistent with each other, and the spectrum of the low-lying $\Omega_c$ states have not been established so far. The situation is even worse for the low-lying $\Omega_b$ spectrum. Compared with the excited $\Omega_c$ states, the strong decays of the low-lying $\Omega_b$ states were only investigated by several works~\cite{Agaev:2017ywp,Wang:2017kfr,Wang:2018fjm,Yao:2018jmc,Santopinto:2018ljf,Chen:2018vuc,Chen:2019ywy}. Nevertheless, due to the lack of experimental information, the choices of parameters may
be adjustable and indeterminate, and the predictions of these different works
do not agree with each other. Hence, it is crucial to investigate the strong decays of these low-lying $\Omega_b$ baryons within the unified model and coincident parameters, which has been adopted to describe the excited $\Lambda_{c(b)}$ and $\Sigma_{c(b)}$ states successfully~\cite{Lu:2018utx,Liang:2019aag,Lu:2019rtg}.

In this work, we calculate the strong decay behaviors of the low-lying $\Omega_b$ states within the $^3 P_0$ model systematically. Our results show that the observed $\Omega_b(6316)^-$, $\Omega_b(6330)^-$, $\Omega_b(6340)^-$, and $\Omega_b(6350)^-$ resonances can be reasonably assigned as the $\lambda-$mode $\Omega_b(1P)$ states with $J^P=1/2^-, 3/2^-, 3/2^-$, and $5/2^-$. Meanwhile, the remaining $P-$wave state with $J^P=1/2^-$ is predicted to be rather broad, which can hardly be observed by experiments. Moreover, the strong decays of the $\Omega_b(2S)$ and $\Omega_b(1D)$ states suggest that these states have relatively narrow total widths and mainly decay into the $\Xi_b \bar K$, $\Xi_b^\prime \bar K$ and $\Xi_b^{\prime*} \bar K$ final states. We hope these abundant theoretical predictions can provide helpful information for the future experimental searches.

This paper is organized as follows. In Sec.~\ref{model}, a brief introduction of the $^3 P_0$ model and notations are illustrated. The strong decays of the low-lying $\Omega_b$ states are presented in Sec.~\ref{low-lying}. A short summary is given in the last section.

\section{$^3P_0$ Model and notations}{\label{model}}
In this work, we adopt the $^3P_0$ model to calculate the two-body OZI-allowed strong decays of the low-lying $\Omega_b$ states. In this model, a quark-antiquark pair with the quantum number $J^{PC}$ =$0^{++}$ is created from the vacuum, and then regroups into two outgoing hadrons by a quark rearrangement process~\cite{micu}. This model has been successfully employed to study different kinds of the hadron systems for their strong decays with considerable successes~\cite{micu,3p0model1,3p0model2,3p0model4,3p0model5,3p0model6,Chen:2007xf,Zhao:2016qmh,Ye:2017dra,Chen:2017gnu,Chen:2016iyi,Lu:2014zua,Lu:2016bbk,Ferretti:2014xqa,Godfrey:2015dva,Segovia:2012cd,Mu:2014iaa,Lu:2018utx,Guo:2019ytq,Liang:2019aag,Lu:2019rtg}. Here, we perform a brief introduction of this model. In the nonrelativistic limit, to describe the decay process $A\rightarrow BC$, the transition operator $T$  in the $^3P_0$ model can be taken as
\begin{eqnarray}
T&=&-3\gamma\sum_m\langle 1m1-m|00\rangle\int
d^3\boldsymbol{p}_4d^3\boldsymbol{p}_5\delta^3(\boldsymbol{p}_4+\boldsymbol{p}_5)\nonumber\\&&\times {\cal{Y}}^m_1\left(\frac{\boldsymbol{p}_4-\boldsymbol{p}_5}{2}\right)\chi^{45}_{1,-m}\phi^{45}_0\omega^{45}_0b^\dagger_{4i}(\boldsymbol{p}_4)d^\dagger_{4j}(\boldsymbol{p}_5),
\end{eqnarray}
where $\gamma$ is a dimensionless $q_4\bar{q}_5$ pair-production strength, and $\boldsymbol{p}_4$ and $\boldsymbol{p}_5$ are the momenta of the created quark $q_4$ and antiquark  $\bar{q}_5$, respectively. The $i$ and $j$ are the color indices of the created quark and antiquark. $\phi^{45}_{0}=(u\bar u + d\bar d +s\bar s)/\sqrt{3}$, $\omega^{45}=\delta_{ij}$, and $\chi_{{1,-m}}^{45}$ are the flavor singlet, color singlet, and spin triplet wave functions of the  $q_4\bar{q}_5$, respectively. The ${\cal{Y}}^m_1(\boldsymbol{p})\equiv|p|Y^m_1(\theta_p, \phi_p)$ is the solid harmonic polynomial reflecting the $P-$wave momentum-space distribution of the created quark pair.

\begin{figure}[!htbp]
\includegraphics[scale=0.6]{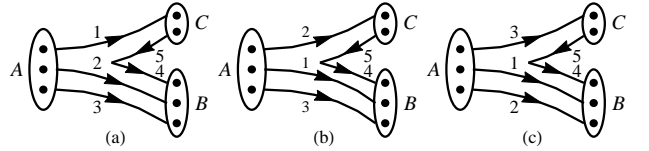}
\vspace{0.0cm} \caption{The baryon decay process $A\to B+C$ in the $^3P_0$ model.}
\label{qpc}
\end{figure}

For the strong decay of a baryon $\Omega_b$, there are three possible rearrangements,
\begin{eqnarray}
A(s_1,s_2,b_3)+P(q_4,\bar q_5)\to B(s_2,q_4,b_3)+C(s_1,\bar q_5),\\
A(s_1,s_2,b_3)+P(q_4,\bar q_5)\to B(s_1,q_4,b_3)+C(s_2,\bar q_5),\\
A(s_1,s_2,b_3)+P(q_4,\bar q_5)\to B(s_1,s_2,q_4)+C(b_3,\bar q_5).
\end{eqnarray}
These three ways of recouplings are also presented in Fig.~\ref{qpc}. It should be mentioned that the
first and second ones stand for the heavy baryon plus
the light meson channels, while the last one denotes the light
baryon plus the heavy meson decay mode.

The $S$ matrix can be written as
\begin{eqnarray}
\langle
f|S|i\rangle=I-i2\pi\delta(E_f-E_i){\cal{M}}^{M_{J_A}M_{J_B}M_{J_C}},
\end{eqnarray}
where the ${\cal{M}}^{M_{J_A}M_{J_B}M_{J_C}}$ is the helicity amplitude of the decay process $A\to B+C$. The explicit expression of the helicity amplitude ${\cal{M}}^{M_{J_A}M_{J_B}M_{J_C}}$ can be found in Refs.~\cite{Chen:2007xf,Lu:2018utx,Liang:2019aag,Lu:2019rtg}.

In this work, we adopt the simplest vertex which assumes a spatially constant pair production strength $\gamma$, the relativistic phase space, and the simple harmonic oscillator wave functions~\cite{micu}. Then, the decay width $\Gamma(A\rightarrow BC)$ is calculated directly
\begin{eqnarray}
\Gamma= \pi^2\frac{p}{M^2_A}\frac{1}{2J_A+1}\sum_{M_{J_A},M_{J_B},M_{J_C}}|{\cal{M}}^{M_{J_A}M_{J_B}M_{J_C}}|^2,
\end{eqnarray}
where $p=|\boldsymbol{p}|=\frac{\sqrt{[M^2_A-(M_B+M_C)^2][M^2_A-(M_B-M_C)^2]}}{2M_A}$,
and $M_A$, $M_B$, and $M_C$ are the masses of the hadrons $A$, $B$,
and $C$, respectively.

For the $\Omega_b(1P)$ states, we employ the masses of four newly observed structures from LHCb experimental data by
assuming that they are possible candidates. For the masses of the $\Omega_b(2S)$ and $\Omega_b(1D)$ states, we adopt the theoretical predictions in the relativistic quark model~\cite{Ebert:2011kk} which are listed in Tab.~\ref{bmass}. For the final ground states, their masses are taken from the Review of Particle Physics~\cite{Tanabashi:2018oc}. For the harmonic oscillator parameters of mesons, we use the effective values as in Ref.~\cite{Godfrey:2015dva}. For the baryon parameters, we use $\alpha_\rho=440~\rm{MeV}$ and
\begin{eqnarray}
\alpha_\lambda=\Bigg(\frac{3m_Q}{2m_q+m_Q} \Bigg)^{1/4} \alpha_\rho,
\end{eqnarray}
where the $m_Q$ and $m_q$ are the heavy and light quark masses, respectively~\cite{Wang:2017hej,Lu:2018utx,Zhong:2007gp,Liang:2019aag,Lu:2019rtg}. The $m_{u/d}=220~\rm{MeV}$, $m_s=419~\rm{MeV}$, $m_c=1628~\rm{MeV}$ and $m_b=4977~\rm{MeV}$ are introduced to consider the mass differences of the heavy and light quarks~\cite{Capstick:1986bm,Godfrey:2015dva,Godfrey:1985xj}. The overall parameter $\gamma$ is determined by the well established $\Sigma_c(2520)^{++} \to \Lambda_c \pi^+$ process. Then, the $\gamma=9.83$ is obtained by reproducing the width $\Gamma [\Sigma_c(2520)^{++} \to \Lambda_c \pi^+]=14.78~\rm{MeV}$~\cite{Tanabashi:2018oc,Lu:2018utx}. With this overall parameter $\gamma$, the strong decay behaviors of the ground and excited $\Lambda_Q$ and $\Sigma_Q$ states can be well described~\cite{Lu:2018utx,Liang:2019aag,Lu:2019rtg}.

\begin{table*}[htb]
\begin{center}
\caption{ \label{bmass} Notations, quantum numbers and masses of initial baryons. The $n_\rho$ and $L_\rho$ denote the nodal quantum number and orbital angular momentum between the two strange quarks, respectively. The $n_\lambda$ and $L_\lambda$ represent the the nodal quantum number and orbital angular momentum between the bottom quark and strange quark system. The $S_\rho$ stands for the total spin of the two strange quarks, $L$ is the total orbital angular momentum, $j$ represent total angular momentum of $L$ and $S_\rho$, $J$ is the total angular momentum, and $P$ is the parity. The masses are taken from the theoretical predictions from relativistic quark model~\cite{Ebert:2011kk}. The units are in MeV. }
\renewcommand\arraystretch{1.8}
\footnotesize
\begin{tabular*}{18cm}{@{\extracolsep{\fill}}*{11}{p{1.5cm}<{\centering}}}
\hline\hline
State      &   $j$	&  $J^P$ & $L$	& $n_{\rho}$ & $n_{\lambda}$ & $L_{\rho}$ & $L_{\lambda}$ & $S_{\rho}$  &Mass\\
\hline
$\Omega_{b}(2S)$	&	1	&	$\frac{1}{2}^+$	&	0	&	0	&	1	&	0	&	0	&	1            &  6450    \\
$\Omega^*_{b}(2S)$	&	1	&	$\frac{3}{2}^+$	&	0	&	0	&	1	&	0	&	0	&	1            &	6461  \\

$\Omega_{b0}(\frac{1}{2}^-)$ &	0	&   $\frac{1}{2}^-$ 	&	1	&	0	&	0	&	0	&	1	&	1     &	6339 \\
$\Omega_{b1}(\frac{1}{2}^-)$	&	1	&	$\frac{1}{2}^-$   &	1	&	0	&	0	&	0	&	1	&	1     &	6330  \\
$\Omega_{b1}(\frac{3}{2}^-)$	&	1	&	$\frac{3}{2}^-$	&	1	&	0	&	0	&	0	&	1	&	1     &	6340 \\
$\Omega_{b2}(\frac{3}{2}^-)$	&	2	&	$\frac{3}{2}^-$	&	1	&	0	&	0	&	0	&	1	&	1     &	6331  \\
$\Omega_{b2}(\frac{5}{2}^-)$	&	2	&   $\frac{5}{2}^-$	&	1	&	0	&	0	&	0	&	1	&	1     &	6334  \\

$\Omega_{b1}(\frac{1}{2}^+)$	&	1	&	$\frac{1}{2}^+$	&	2	&	0	&	0	&	0	&	2	&	1     & 6540 \\
$\Omega_{b1}(\frac{3}{2}^+)$	&	1	&	$\frac{3}{2}^+$	&	2	&	0	&	0	&	0	&	2	&	1     & 6549  \\
$\Omega_{b2}(\frac{3}{2}^+)$	&	2	&	$\frac{3}{2}^+$	&	2	&	0	&	0	&	0	&	2	&	1     &	6530   \\
$\Omega_{b2}(\frac{5}{2}^+)$	&	2	&	$\frac{5}{2}^+$   &	2	&	0	&	0	&	0	&	2	&	1     & 6529 \\
$\Omega_{b3}(\frac{5}{2}^+)$	&	3	&	$\frac{5}{2}^+$	&	2	&	0	&	0	&	0	&	2	&	1     & 6520	\\
$\Omega_{b3}(\frac{7}{2}^+)$	&	3	&	$\frac{7}{2}^+$   &	2	&	0	&	0	&	0	&	2	&	1     & 6517  \\

\hline\hline
\end{tabular*}
\end{center}
\end{table*}

\section{Strong decay}{\label{low-lying}}

\subsection{$\Omega_b(1P)$ states}
There are five $\lambda$-mode $\Omega_b(1P)$ states in the constituent quark model, which are denoted as $\Omega_{b0}(\frac{1}{2}^-)$, $\Omega_{b1}(\frac{1}{2}^-)$, $\Omega_{b1}(\frac{3}{2}^-)$, $\Omega_{b2}(\frac{3}{2}^-)$ and $\Omega_{b2}(\frac{5}{2}^-)$, respectively. From Tab.~\ref{bmass}, the predicted masses of the five $\Omega_b(1P)$ states are around $6330\sim 6340$ MeV. As mentioned in the Introduction, four narrow structures $\Omega_b(6316)^-$, $\Omega_b(6330)^-$, $\Omega_b(6340)^-$, and $\Omega_b(6350)^-$ have been observed in the $\Xi_b^0 K^-$ mass spectrum. Given the uncertainties of quark model prediction, these structures are good candidates of the $\Omega_b(1P)$ states. With these experimental masses, all possibilities of these four observed resonances as the $\Omega_b(1P)$ states are considered and the total decay widths are listed in the Tab.~\ref{B1Pexperiment}. For the $j=0$ state, the total decay width is predicted to be rather large, while for the two $j=1$ states, their OZI-allowed strong decays are forbidden due to the quantum number conservation and the phase space constraint. For the two $j=2$ states, the total decay widths are about several MeV, which may correspond to the experimental observations. It can be seen that the pure $\Omega_b(1P)$ assignments can hardly interpret these four resonances simultaneously.

\begin{table*}
\begin{center}
\caption{\label{B1Pexperiment}The decay widths of the four observed resonances as the $\Omega_b(1P)$ states in MeV.}
\renewcommand{\arraystretch}{1.5}
\begin{tabular*}{18cm}{@{\extracolsep{\fill}}*{11}{p{1.8cm}<{\centering}}}
\hline\hline
 &	$\Omega_{b0}(\frac{1}{2}^-,1P)$	&	$\Omega_{b1}(\frac{1}{2}^-,1P)$	&	$\Omega_{b1}(\frac{3}{2}^-,1P)$	&	$\Omega_{b2}(\frac{3}{2}^-,1P)$	&	$\Omega_{b2}(\frac{5}{2}^-,1P)$	&Experiments \\
$\Omega_{b}(6316)$	&	870.52	&	$-$	&	$-$	&	0.35	&	0.35 & $<2.8$	\\
$\Omega_{b}(6330)$	&	1056.79	&	$-$	&	$-$	&	1.08	&	1.08 &	$<3.1$\\
$\Omega_{b}(6340)$	&	1146.35	&	$-$	&	$-$	&	1.85	&	1.85 &	$<1.5$\\
$\Omega_{b}(6350)$	&	1224.29	&	$-$ &	$-$	&	2.98	&	2.98 &	 $1.4^{+1.0}_{-0.8}\pm0.1$\\
\hline\hline
\end{tabular*}
\end{center}
\end{table*}

\begin{figure}[!htbp]
\includegraphics[scale=0.27]{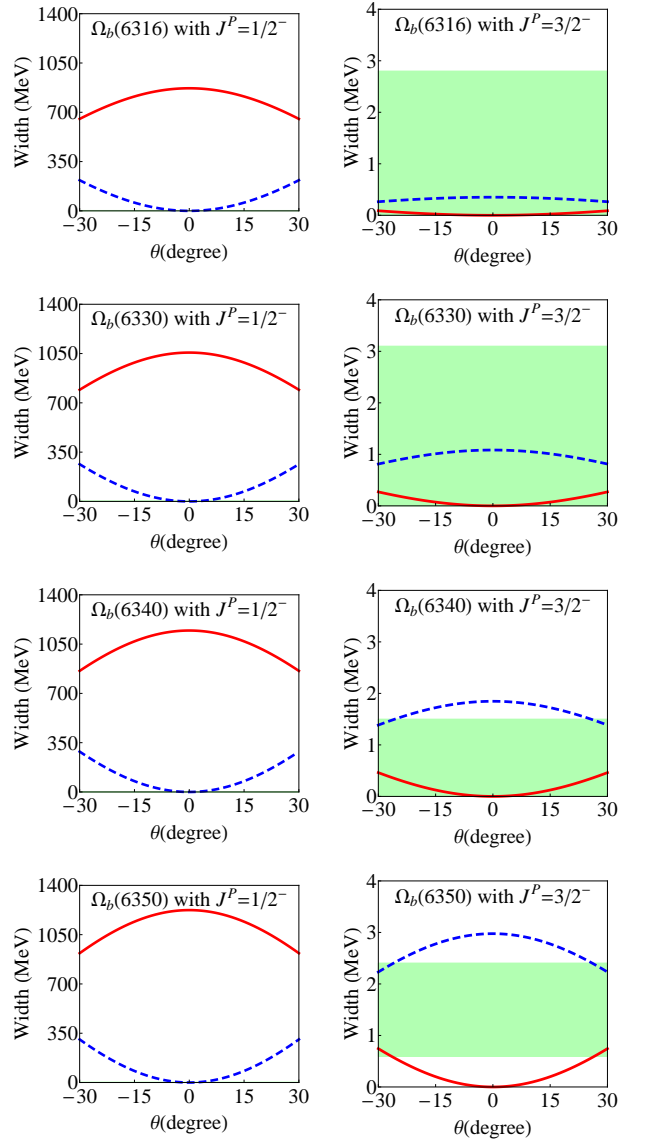}
\vspace{0.0cm} \caption{The total decay widths of various assignments as functions of the mixing angle $\theta$ in the range $-30^\circ \sim 30^\circ$. The red solid lines are the $|1P~{1/2^-}\rangle_1$ states, and the blue dashed curves correspond to the $|1P~{1/2^-}\rangle_2$ states. The green bands stand for the measured total decay widths with errors.}
\label{omegab1pmix}
\end{figure}

In fact, the physical resonances can be the mixing of the quark model states with the same $J^P$, that is
\begin{equation}
\left(\begin{array}{c}| 1P~{1/2^-}\rangle_1\cr |  1P~{1/2^-}\rangle_2
\end{array}\right)=\left(\begin{array}{cc} \cos\theta & \sin\theta \cr -\sin\theta &\cos\theta
\end{array}\right)
\left(\begin{array}{c} |1/2^-,j=0
\rangle \cr |1/2^-,j=1\rangle
\end{array}\right),
\end{equation}
\begin{equation}
\left(\begin{array}{c}| 1P~{3/2^-}\rangle_1\cr |  1P~{3/2^-}\rangle_2
\end{array}\right)=\left(\begin{array}{cc} \cos\theta & \sin\theta \cr -\sin\theta &\cos\theta
\end{array}\right)
\left(\begin{array}{c} |3/2^-,j=1
\rangle \cr |3/2^-,j=2\rangle
\end{array}\right).
\end{equation}
Under the heavy quark limit, the mixing angle should equals to zero. Given the finite mass of the bottom quark, the heavy quark symmetry should be approximately preserved, and the mixing angle between the physical states and the quark model states can have a small divergence with the $j-j$ coupling scheme. The total decay widths of various assignments versus the mixing angle $\theta$ in the range $-30^\circ \sim 30^\circ$ are plotted in Fig.~\ref{omegab1pmix}.

With the similar masses and total widths, the strong decays of the four resonances are all consistent with the experimental data under the $J^P=1/2^-$ and $J^P=3/2^-$ assignments. However, only three states $| 1P~{1/2^-}\rangle_2 $, $| 1P~{3/2^-}\rangle_1 $, and $| 1P~{3/2^-}\rangle_2 $ belong to the narrow resonances, while the $| 1P~{1/2^-}\rangle_1 $ state is too large to be occupied. Fortunately, there is also a $J^P=5/2^-$ state, which is suitable for all four resonances. Hence, four narrow structures $\Omega_b(6316)^-$, $\Omega_b(6330)^-$, $\Omega_b(6340)^-$, and $\Omega_b(6350)^-$ can be clarified into the $| 1P~{1/2^-}\rangle_2 $, $| 1P~{3/2^-}\rangle_1 $, $| 1P~{3/2^-}\rangle_2 $, and $| 1P~{5/2^-}\rangle $ states. However, from current experimental information, we can hardly provide the exact correspondence between the physical resonances and theoretical states.

The predicted total decay width of the $| 1P~{1/2^-}\rangle_1 $ state is rather large, which can be hardly observed experimentally. This explains why the LHCb experiment only observed four peaks in the $\Xi_b^0 K^-$ mass spectrum. Our present results are consistent with the previous chiral quark model calculations under the $j-j$ coupling scheme~\cite{Wang:2018fjm}. In Ref.~\cite{Chen:2018vuc}, the authors predicted one narrow $j=0$ state, two OZI-forbidden $j=1$ states, and two narrow $j=2$ states with the $j-j$ couplings. If the mixing mechanism was considered, the authors could also perform four narrow $\Omega_b(1P)$ states which agree with our present calculations except for the $j=0$ state. Moreover, based on the $L-S$ couplings, the authors predicted five narrow $\Omega_b(1P)$ states within the $^3P_0$ model~\cite{Santopinto:2018ljf}, and three narrow states within the chiral quark model~\cite{Wang:2017kfr}, which show different features with our results. Present experiment data suggests that the physical states of these $P-$wave $\Omega_b$ baryons more favor the $j-j$ coupling scheme.

\begin{figure}[!htbp]
\includegraphics[scale=0.75]{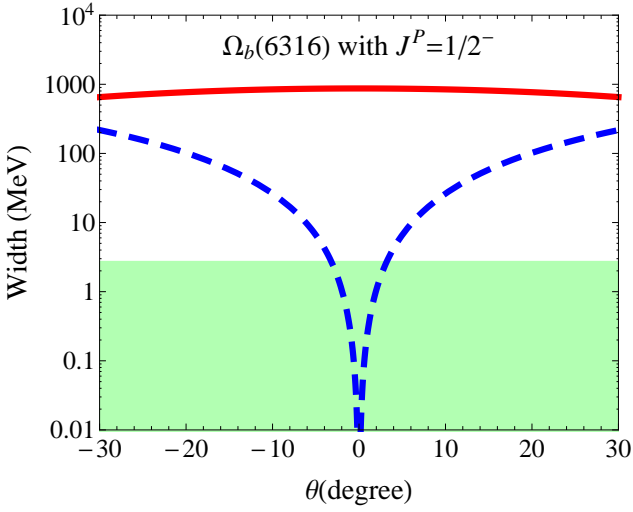}
\vspace{0.0cm} \caption{The allowed mixing angle $\theta$ under the assignment of the $\Omega_b(6316)^-$ as $J^P=1/2^-$ state. The red solid line is the $|1P~{1/2^-}\rangle_1$ state, and the blue dashed curve corresponds to the $|1P~{1/2^-}\rangle_2$ state. The green band stands for the measured total decay width with error.}
\label{mix}
\end{figure}

Also, one can determine the mixing angle $\theta$ of the two $J^P=1/2^-$ states. If we choose the normal mass order for the $\Omega_b(1P)$ states, the mixing angle $\theta$ can be constrained by the width of $\Omega_b(6316)^-$ resonance. From Fig.~\ref{mix}, one can see that the mixing angle $\theta$ should lie in $-3.3^\circ \sim 3.3^\circ$ except for zero. When the mixing angle equals to zero, the $| 1P~{1/2^-}\rangle_2 $ state corresponds to the $\Omega_{b1}(\frac{1}{2}^-)$, which cannot decay into the $\Xi_b \bar K$ channel. Thanks to the finite mass of bottom quark, the small mixing between two $J^P=1/2^-$ states can occur, and the narrow one can be observed in the $\Xi_b \bar K$ final state experimentally. For the two $J^P=3/2^-$ states, the mixing angle cannot be determined by current experimental data.  More theoretical and experimental efforts are needed to solve this problem.

\subsection{$\Omega_b(2S)$ states}
In the traditional quark model, there are two $\lambda$-mode $2S-$wave excitations, which are denoted as $\Omega_{b}(2S)$ and $\Omega^*_{b}(2S)$. From Tab.~\ref{bmass}, the predicted masses of these two $\Omega_{b}(2S)$ states are 6450 MeV and 6461 MeV, respectively. With the calculated masses, their strong decays within the $^3{P_0}$ model are estimated and shown in Tab.~\ref{B2S}.

The total decay width of the $J^P=1/2^+$ state is about 50 MeV and the dominating decay mode is $\Xi_b \bar K$. The branching ratios are predicted to be
\begin{equation}
Br(\Xi^0_b K^-,\Xi^-_b \bar K^0) \thicksim 48\%,45\%.
\end{equation}
For the $J^P=3/2^+$ state, the total width is predicted to be 53 MeV and the $\Xi_b \bar K$ channel also dominates. The branching ratios of the dominating channels $\Xi^0_b \bar K^0$ and $\Xi^+_b K^-$ are
\begin{equation}
Br(\Xi^0_b K^-,\Xi^-_b \bar K^0)\thicksim 48\%,46\%.
\end{equation}
The total decay widths of our predictions are larger than that of the potential model, but the main decay mode is consistent with each other~\cite{Chen:2018vuc}. With the relatively narrow total widths and dominating decay mode $\Xi_b \bar K$, these two $\Omega_b(2S)$ states have good potentials to be observed in the $\Xi_b \bar K$ mass spectrum in future experiments.

\begin{table}[!htbp]
\begin{center}
\caption{\label{B2S}Decay widths of the $\Omega_b(2S)$ states in MeV.}
\renewcommand{\arraystretch}{1.5}
\normalsize
\begin{tabular*}{8.5cm}{@{\extracolsep{\fill}}*{3}{p{2.5cm}<{\centering}}}
\hline\hline
Mode	&	$\Omega_b(2S )$	&	$\Omega^*_b(2S )$	\\			
$\Xi^0_b K^-$	        &	23.76 	&	25.47 	\\
$\Xi^-_b \bar K^0$	    &	22.56 	&	24.32 	\\
$\Xi^{'0}_b K^-$	    &	1.87 	&	0.86 	\\
$\Xi^{'-}_b \bar K^0$	&	1.40 	&	0.72 	\\
$\Xi^{'*0}_b K^-$	    &	0.08 	&	1.39 	\\
$\Xi^{'*-}_b \bar K^0$	&	$-$   	&	0.56 	\\
Total width	            &	49.67 	&	53.32 	\\	
\hline\hline
\end{tabular*}
\end{center}
\end{table}

\subsection{$\Omega_b(1D)$ states}

From Tab.~\ref{bmass}, the masses of the six $\Omega_b(1D)$ states are predicted to be around $6517\sim 6549$ MeV. The strong decay widths for these $D-$wave states are estimated and presented in Tab.~\ref{B1D}. It can be seen that the total decay widths of $\Omega_{b1}(\frac{1}{2}^+)$, $\Omega_{b1}(\frac{3}{2}^+)$, $\Omega_{b2}(\frac{3}{2}^+)$, $\Omega_{b2}(\frac{5}{2}^+)$, $\Omega_{b3}(\frac{5}{2}^+)$ and $\Omega_{b3}(\frac{7}{2}^+)$ states are about 106, 108, 27, 21, 3, and 3 MeV, respectively. For the two $j=1$ states, the dominating decay mode is $\Xi_b \bar K$, while other decay channels are relatively small. For the two $j=2$ states, the $\Xi^0_b \bar K^0$ and $\Xi^+_b K^-$ final states are forbidden due to the quantum number conservation, and the main decay modes for the $\Omega_{b2}(\frac{3}{2}^+)$ and $\Omega_{b2}(\frac{5}{2}^+)$ states are $\Xi^{'}_b \bar K$ and $\Xi^{'*}_b \bar K$, respectively. To distinguish these two $j=2$ states, the partial decay ratios between $\Xi^{'}_b \bar K$ and $\Xi^{'*}_b \bar K$ modes are helpful. For the two $j=3$ states, the calculated decay widths are about several MeV, which are quite narrow. The main decay mode is $\Xi_b \bar K$ with the branching ratio up to 97.3\% and 98.2\% for the $J^P=5/2^+$ and $J^P=7/2^+$ states, respectively. 

Within the chiral quark model and the $L-S$ scheme, the authors predicted that the total decay widths of these six states lie in $7\sim 29$ MeV~\cite{Yao:2018jmc}, which show different features with ours. Also, our predictions on these $D-$wave $\Omega_b$ states are quite different with the potential model calculations, where six $\Omega_b(1D)$ states all have widths of several MeV~\cite{Chen:2019ywy}. Future experimental searches can help us to clarify these theoretical divergences.

\begin{table*}
\begin{center}
\caption{ \label{B1D}Decay widths of the $\Omega_b(1D)$ states in MeV.}
\renewcommand{\arraystretch}{1.5}
\begin{tabular*}{18cm}{@{\extracolsep{\fill}}*{11}{p{1.5cm}<{\centering}}}
\hline\hline
Mode	&	$\Omega_{b1}(\frac{1}{2}^+)$	&	$\Omega_{b1}(\frac{3}{2}^+)$	&	$\Omega_{b2}(\frac{3}{2}^+)$	&	$\Omega_{b2}(\frac{5}{2}^+)$	&	$\Omega_{b3}(\frac{5}{2}^+)$	&	$\Omega_{b3}(\frac{7}{2}^+)$	\\
$\Xi^0_b K^-$	        &	45.29 	&	46.45 	&	$-$	    &	$-$	    &	1.54 	&	1.46 	\\
$\Xi^-_b \bar K^0$	    &	44.37 	&	45.61 	&	$-$	    &	$-$	    &	1.36 	&	1.29 	\\
$\Xi^{'0}_b K^-$	    &	5.99 	&	1.66 	&	11.87 	&	0.03 	&	0.03 	&	0.01 	\\
$\Xi^{'-}_b \bar K^0$	&	5.79 	&	1.61 	&	11.40 	&	0.03 	&	0.02 	&	0.01 	\\
$\Xi^{'*0}_b K^-$	    &	2.38 	&	6.75 	&	1.86 	&	10.82 	&	0.02 	&	0.02 	\\
$\Xi^{'*-}_b \bar K^0$	&	2.17 	&	6.22 	&	1.67 	&	9.70 	&	0.01 	&	0.01 	\\
Total width	            &	105.99 	&	108.30 	&	26.80 	&	20.58 	&	2.98 	&	2.80 	\\

\hline\hline
\end{tabular*}
\end{center}
\end{table*}

\section{summary}{\label{summary}}
In this work, we study the strong decay behaviors of the low-lying $\lambda-$mode $\Omega_{b}$ states within the $^3P_0$ model systematically. According to the masses and decay modes, four newly observed $\Omega_{b}$ resonances can be reasonably interpreted. It is found that the $\Omega_b(6316)^-$, $\Omega_b(6330)^-$, $\Omega_b(6340)^-$, and $\Omega_b(6350)^-$ can be assigned as the $\lambda-$mode $\Omega_b(1P)$ states with $J^P=1/2^-, 3/2^-, 3/2^-$, and $5/2^-$. Meanwhile, the remaining  $P-$wave state with $J^P=1/2^-$ should have a rather broad width, which can hardly be observed by experiments. For the $\Omega_b(2S)$ and $\Omega_b(1D)$ states, our predictions suggest that these states have relatively narrow total widths and mainly decay into the $\Xi_b \bar K$, $\Xi_b^\prime \bar K$ and $\Xi_b^{\prime*} \bar K$ final states. We hope these theoretical calculations are valuable for searching more excited $\Omega_b$ states in future experiments.

\bigskip
\noindent
\begin{center}
{\bf ACKNOWLEDGEMENTS}\\

\end{center}
We would like to thank Xian-Hui Zhong and Long-Cheng Gui for valuable discussions. This project is supported by the National Natural Science Foundation of China under Grants No. 11705056, No.~11775078, and No.~U1832173.


\begin{thebibliography}{99}
\bibitem{Chen:2016spr}
  H.~X.~Chen, W.~Chen, X.~Liu, Y.~R.~Liu and S.~L.~Zhu,
  A review of the open charm and open bottom systems,
  Rept.\ Prog.\ Phys.\  {\bf 80}, 076201 (2017).

\bibitem{Cheng:2015iom}
  H.~Y.~Cheng,
  Charmed baryons circa 2015,
  Front.\ Phys.\ (Beijing) {\bf 10}, 101406 (2015).

\bibitem{Crede:2013sze}
  V.~Crede and W.~Roberts,
  Progress towards understanding baryon resonances,
  Rept.\ Prog.\ Phys.\  {\bf 76}, 076301 (2013).

\bibitem{Klempt:2009pi}
  E.~Klempt and J.~M.~Richard,
  Baryon spectroscopy,
  Rev.\ Mod.\ Phys.\  {\bf 82}, 1095 (2010).


\bibitem{Tanabashi:2018oc}
  M.~Tanabashi {\it et al.} (Particle Data Group),
  Phys.\ Rev.\ D {\bf 98}, 030001 (2018).


\bibitem{Aaij:2017nav}
  R.~Aaij {\it et al.} (LHCb Collaboration),
  Observation of five new narrow $\Omega_c^0$ states decaying to $\Xi_c^+ K^-$,
  Phys.\ Rev.\ Lett.\  {\bf 118}, 182001 (2017).

\bibitem{Yelton:2017qxg}
  J.~Yelton {\it et al.} (Belle Collaboration),
  Observation of Excited $\Omega_c$ Charmed Baryons in $e^+e^-$ Collisions,
  Phys.\ Rev.\ D {\bf 97}, 051102 (2018).

\bibitem{Aaij:2020cex}
  R.~Aaij {\it et al.} (LHCb Collaboration),
  First observation of excited $\Omega_b^-$ states,
  arXiv:2001.00851.

\bibitem{Ebert:2011kk}
  D.~Ebert, R.~N.~Faustov and V.~O.~Galkin,
  Spectroscopy and Regge trajectories of heavy baryons in the relativistic quark-diquark picture,
  Phys.\ Rev.\ D {\bf 84}, 014025 (2011).

\bibitem{Roberts:2007ni}
  W.~Roberts and M.~Pervin,
  Heavy baryons in a quark model,
  Int.\ J.\ Mod.\ Phys.\ A {\bf 23}, 2817 (2008).

\bibitem{Ebert:2007nw}
  D.~Ebert, R.~N.~Faustov and V.~O.~Galkin,
  Masses of excited heavy baryons in the relativistic quark model,
  Phys.\ Lett.\ B {\bf 659}, 612 (2008).

\bibitem{Garcilazo:2007eh}
  H.~Garcilazo, J.~Vijande and A.~Valcarce,
  Faddeev study of heavy baryon spectroscopy,
  J.\ Phys.\ G {\bf 34}, 961 (2007).

\bibitem{Yoshida:2015tia}
  T.~Yoshida, E.~Hiyama, A.~Hosaka, M.~Oka and K.~Sadato,
  Spectrum of heavy baryons in the quark model,
  Phys.\ Rev.\ D {\bf 92}, 114029 (2015).

\bibitem{Thakkar:2016dna}
  K.~Thakkar, Z.~Shah, A.~K.~Rai and P.~C.~Vinodkumar,
  Excited State Mass spectra and Regge trajectories of Bottom Baryons,
  Nucl.\ Phys.\ A {\bf 965}, 57 (2017).
  
\bibitem{Chen:2018vuc}
  B.~Chen and X.~Liu,
  Assigning the newly reported $\Sigma_b(6097)$ as a $P$-wave excited state and predicting its partners,
  Phys.\ Rev.\ D {\bf 98}, 074032 (2018).

\bibitem{Shah:2018daf}
  Z.~Shah and A.~K.~Rai,
  Mass Spectra of Singly Beauty $\Omega_b^{-}$ Baryon,
  Few Body Syst.\  {\bf 59}, 112 (2018).

\bibitem{Qin:2019hgk}
  S.~X.~Qin, C.~D.~Roberts and S.~M.~Schmidt,
  Spectrum of light- and heavy-baryons,
  Few Body Syst.\  {\bf 60}, 26 (2019).


\bibitem{Chen:2017sci}
  H.~X.~Chen, Q.~Mao, W.~Chen, A.~Hosaka, X.~Liu and S.~L.~Zhu,
  Decay properties of $P$-wave charmed baryons from light-cone QCD sum rules,
  Phys.\ Rev.\ D {\bf 95}, 094008 (2017).
  


\bibitem{Karliner:2017kfm}
  M.~Karliner and J.~L.~Rosner,
  Very narrow excited $\Omega_c$ baryons,
  Phys.\ Rev.\ D {\bf 95}, 114012 (2017).

\bibitem{Wang:2017hej}
  K.~L.~Wang, L.~Y.~Xiao, X.~H.~Zhong and Q.~Zhao,
  Understanding the newly observed $\Omega_c$ states through their decays,
  Phys.\ Rev.\ D {\bf 95}, 116010 (2017).

\bibitem{Wang:2017vnc}
  W.~Wang and R.~L.~Zhu,
  Interpretation of the newly observed $\Omega_c^0$ resonances,
  Phys.\ Rev.\ D {\bf 96}, 014024 (2017).

\bibitem{Padmanath:2017lng}
  M.~Padmanath and N.~Mathur,
  Quantum Numbers of Recently Discovered $\Omega^{0}_{c}$ Baryons from Lattice QCD,
  Phys.\ Rev.\ Lett.\  {\bf 119}, 042001 (2017).

\bibitem{Cheng:2017ove}
  H.~Y.~Cheng and C.~W.~Chiang,
  Quantum numbers of $\Omega_c$ states and other charmed baryons,
  Phys.\ Rev.\ D {\bf 95}, 094018 (2017).

\bibitem{Huang:2017dwn}
  H.~Huang, J.~Ping and F.~Wang,
  Investigating the excited $\Omega^{0}_{c}$ states through $\Xi_{c}K$ and $\Xi^{'}_{c}K$ decay channels,
  Phys.\ Rev.\ D {\bf 97}, 034027 (2018).

\bibitem{Wang:2017zjw}
  Z.~G.~Wang,
  Analysis of $\Omega _c(3000)$ , $\Omega _c(3050)$ , $\Omega _c(3066)$ , $\Omega _c(3090)$ and $\Omega _c(3119)$ with QCD sum rules,
  Eur.\ Phys.\ J.\ C {\bf 77}, 325 (2017).

\bibitem{Zhao:2017fov}
  Z.~Zhao, D.~D.~Ye and A.~Zhang,
  Hadronic decay properties of newly observed $\Omega_c$ baryons,
  Phys.\ Rev.\ D {\bf 95}, 114024 (2017).

\bibitem{Chen:2017gnu}
  B.~Chen and X.~Liu,
  New $\Omega_c^0$ baryons discovered by LHCb as the members of $1P$ and $2S$ states,
  Phys.\ Rev.\ D {\bf 96}, 094015 (2017).


\bibitem{Kim:2017jpx}
  H.~C.~Kim, M.~V.~Polyakov and M.~Praszalowicz,
  Possibility of the existence of charmed exotica,
  Phys.\ Rev.\ D {\bf 96}, 014009 (2017).

\bibitem{Agaev:2017lip}
  S.~S.~Agaev, K.~Azizi and H.~Sundu,
  Interpretation of the new $\Omega_c^{0}$ states via their mass and width,
  Eur.\ Phys.\ J.\ C {\bf 77}, 395 (2017).

\bibitem{Wang:2017xam}
  Z.~G.~Wang, X.~N.~Wei and Z.~H.~Yan,
  Revisit assignments of the new excited $\Omega _c$ states with QCD sum rules,
  Eur.\ Phys.\ J.\ C {\bf 77}, 832 (2017).




\bibitem{Debastiani:2017ewu}
  V.~R.~Debastiani, J.~M.~Dias, W.~H.~Liang and E.~Oset,
  Molecular $\Omega_c$ states generated from coupled meson-baryon channels,
  Phys.\ Rev.\ D {\bf 97}, 094035 (2018).

\bibitem{Wang:2017kfr}
  K.~L.~Wang, Y.~X.~Yao, X.~H.~Zhong and Q.~Zhao,
  Strong and radiative decays of the low-lying $S$- and $P$-wave singly heavy baryons,
  Phys.\ Rev.\ D {\bf 96}, 116016 (2017).

\bibitem{Agaev:2017ywp}
  S.~S.~Agaev, K.~Azizi and H.~Sundu,
  Phys.\ Rev.\ D {\bf 96}, 094011 (2017).

\bibitem{Yao:2018jmc}
  Y.~X.~Yao, K.~L.~Wang and X.~H.~Zhong,
  Strong and radiative decays of the low-lying $D$-wave singly heavy baryons,
  Phys.\ Rev.\ D {\bf 98}, 076015 (2018).

\bibitem{Santopinto:2018ljf}
  E.~Santopinto, A.~Giachino, J.~Ferretti, H.~Garcia-Tecocoatzi, M.~A.~Bedolla, R.~Bijker and E.~Ortiz-Pacheco,
  The $\Omega_{c}$-puzzle solved by means of spectrum and strong decay amplitude predictions,
  Eur.\ Phys.\ J.\ C {\bf 79}, 1012 (2019).

\bibitem{Yang:2017rpg}
  G.~Yang and J.~Ping,
  Dynamical study of $\Omega_c^0$ in the chiral quark model,
  Phys.\ Rev.\ D {\bf 97}, 034023 (2018).

\bibitem{An:2017lwg}
  C.~S.~An and H.~Chen,
  Observed $\Omega_{c}^{0}$ resonances as pentaquark states,
  Phys.\ Rev.\ D {\bf 96}, 034012 (2017).


\bibitem{Montana:2017kjw}
  G.~Monta\~na, A.~Feijoo and \`A.~Ramos,
  A meson-baryon molecular interpretation for some $\Omega_{c}$ excited states,
  Eur.\ Phys.\ J.\ A {\bf 54}, 64 (2018).

\bibitem{Wang:2017smo}
  C.~Wang, L.~L.~Liu, X.~W.~Kang, X.~H.~Guo and R.~W.~Wang,
  Possible open-charmed pentaquark molecule $\Omega_c(3188)$ - the $D \Xi$ bound state - in the Bethe-Salpeter formalism,
  Eur.\ Phys.\ J.\ C {\bf 78}, 407 (2018).

\bibitem{Huang:2018wgr}
  Y.~Huang, C.~J.~Xiao, Q.~F.~L\"u, R.~Wang, J.~He and L.~Geng,
  Strong and radiative decays of $D\Xi$ molecular state and newly observed $\Omega_c$ states,
  Phys.\ Rev.\ D {\bf 97}, 094013 (2018).

\bibitem{Debastiani:2018adr}
  V.~R.~Debastiani, J.~M.~Dias, W.~H.~Liang and E.~Oset,
  $\Omega_b^- \to (\Xi_c^+ \, K^-) \, \pi^-$ and the $\Omega_c$ states,
  Phys.\ Rev.\ D {\bf 98}, 094022 (2018).

\bibitem{Wang:2018alb}
  Z.~G.~Wang and J.~X.~Zhang,
  Possible pentaquark candidates: new excited $\Omega _c$ states,
  Eur.\ Phys.\ J.\ C {\bf 78}, 503 (2018).
  


\bibitem{Wang:2018fjm}
  K.~L.~Wang, Q.~F.~L\"u and X.~H.~Zhong,
  Interpretation of the newly observed $\Sigma_b(6097)^{\pm}$ and $\Xi_b(6227)^-$ states as the $P$-wave bottom baryons,
  Phys.\ Rev.\ D {\bf 99}, 014011 (2019)
  
\bibitem{Chen:2019ywy}
  B.~Chen, S.~Q.~Luo, X.~Liu and T.~Matsuki,
  Interpretation of the observed $\Lambda_b(6146)^0$ and $\Lambda_b(6152)^0$ states as 1$D$ bottom baryons,
  Phys.\ Rev.\ D {\bf 100}, 094032 (2019).

\bibitem{Lu:2018utx}
  Q.~F.~L\"u, L.~Y.~Xiao, Z.~Y.~Wang and X.~H.~Zhong,
  Strong decay of $\Lambda _c(2940)$ as a $2P$ state in the $\Lambda _c$ family,
  Eur.\ Phys.\ J.\ C {\bf 78}, 599 (2018).

\bibitem{Liang:2019aag}
  W.~Liang, Q.~F.~L\"u and X.~H.~Zhong,
  Canonical interpretation of the newly observed $\Lambda_b(6146)^0$ and $\Lambda_b(6152)^0$ via strong decay behaviors,
  Phys.\ Rev.\ D {\bf 100}, 054013 (2019).

\bibitem{Lu:2019rtg}
  Q.~F.~L\"u and X.~H.~Zhong,
  Strong decays of the higher excited $\Lambda_Q$ and $\Sigma_Q$ baryons,
  arXiv:1910.06126.

     \bibitem{micu}
L. Micu, Decay rates of meson resonances in a quark model, Nucl. Phys. B {\bf 10}, 521, (1969).






  \bibitem{3p0model1}
A. Le Yaounc, L. Oliver, O. Pene, and J. C. Raynal,
{\sl Hardon Transitons in the quark model} (Gordon and Breach, New York, 1988).

\bibitem{3p0model2}
W. Roberts and B. Silverstr-Brac, General method of calculation of any hadronic decay in the $^3P_0$ model, Few-Body Syst. {\bf 11}, 171 (1992).

\bibitem{3p0model4}
E. S. Ackleh, T. Barnes, and E. S. Swanson, On the mechanism of open flavor strong decays, Phys. Rev. D {\bf 54}, 6811 (1996).

\bibitem{3p0model5}
T. Barnes, F. E. Close, P. R. Page, and E. S. Swanson, Higher quarkonia, Phys. Rev. D {\bf 55}, 4157 (1997).

\bibitem{3p0model6}
T. Barnes, N. Black, and P. R. Page, Strong decays of strange quarkonia, Phys. Rev. D  {\bf 68}, 054014 (2003).


\bibitem{Chen:2007xf}
  C.~Chen, X.~L.~Chen, X.~Liu, W.~Z.~Deng and S.~L.~Zhu,
  Strong decays of charmed baryons,
  Phys.\ Rev.\ D {\bf 75}, 094017 (2007).

\bibitem{Zhao:2016qmh}
  Z.~Zhao, D.~D.~Ye and A.~Zhang,
  Nature of charmed strange baryons $\Xi_c(3055)$ and $\Xi_c(3080)$,
  Phys.\ Rev.\ D {\bf 94}, 114020 (2016).

\bibitem{Ye:2017dra}
  D.~D.~Ye, Z.~Zhao and A.~Zhang,
  Study of $2S$- and $1D$- excitations of observed charmed strange baryons,
  Phys.\ Rev.\ D {\bf 96}, 114003 (2017).

\bibitem{Chen:2016iyi}
  B.~Chen, K.~W.~Wei, X.~Liu and T.~Matsuki,
  Low-lying charmed and charmed-strange baryon states,
  Eur.\ Phys.\ J.\ C {\bf 77}, 154 (2017).

\bibitem{Lu:2014zua}
  Q.~F.~L\"u and D.~M.~Li,
  Understanding the charmed states recently observed by the LHCb and BaBar Collaborations in the quark model,
  Phys.\ Rev.\ D {\bf 90}, 054024 (2014).

\bibitem{Lu:2016bbk}
  Q.~F.~L\"u, T.~T.~Pan, Y.~Y.~Wang, E.~Wang and D.~M.~Li,
  Excited bottom and bottom-strange mesons in the quark model,
  Phys.\ Rev.\ D {\bf 94}, 074012 (2016).

\bibitem{Ferretti:2014xqa}
  J.~Ferretti, G.~Galata and E.~Santopinto,
  Quark structure of the $X(3872)$ and $\chi_b(3P)$ resonances,
  Phys.\ Rev.\ D {\bf 90}, 054010 (2014).

\bibitem{Godfrey:2015dva}
  S.~Godfrey and K.~Moats,
  Properties of Excited Charm and Charm-Strange Mesons,
  Phys.\ Rev.\ D {\bf 93}, 034035 (2016).

\bibitem{Segovia:2012cd}
  J.~Segovia, D.~R.~Entem and F.~Fernandez, Scaling of the $^3P_0$ Strength in Heavy Meson Strong Decays, Phys.\ Lett.\ B {\bf 715}, 322 (2012).

\bibitem{Mu:2014iaa}
  C.~Mu, X.~Wang, X.~L.~Chen, X.~Liu and S.~L.~Zhu,
  Dipion decays of heavy baryons,
  Chin.\ Phys.\ C {\bf 38}, 113101 (2014).



\bibitem{Guo:2019ytq}
  J.~J.~Guo, P.~Yang and A.~Zhang,
  Strong decays of observed $\Lambda_c$ baryons in the $^3P_0$ model,
  Phys.\ Rev.\ D {\bf 100}, 014001 (2019).


\bibitem{Zhong:2007gp}
  X.~H.~Zhong and Q.~Zhao,
  Charmed baryon strong decays in a chiral quark model,
  Phys.\ Rev.\ D {\bf 77}, 074008 (2008).

\bibitem{Godfrey:1985xj}
  S.~Godfrey and N.~Isgur,
  Mesons in a Relativized Quark Model with Chromodynamics,
  Phys.\ Rev.\ D {\bf 32}, 189 (1985).

\bibitem{Capstick:1986bm}
  S.~Capstick and N.~Isgur,
  Baryons in a Relativized Quark Model with Chromodynamics,
  Phys.\ Rev.\ D {\bf 34}, 2809 (1986).




\end{thebibliography}
\end{document}